# Auricular Vagus Nerve Stimulation for Enhancing Remote Pilot Training and Operations.

**William J. Tyler**[1,2]

[1]Department of Biomedical Engineering and
Center for Neuroengineering and Brain Computer Interfaces
University of Alabama at Birmingham, Alabama, 35294
[2]IST, LLC, Birmingham, Alabama, 35242 USA

## Abstract

The rapid growth of the drone industry, particularly in the use of small unmanned aerial systems (sUAS) and unmanned aerial vehicles (UAVs), requires the development of advanced training protocols for remote pilots. Remote pilots must develop a combination of technical and cognitive skills to manage the complexities of modern drone operations. This paper explores the integration of neurotechnology, specifically auricular vagus nerve stimulation (aVNS), as a method to enhance remote pilot training and performance. The scientific literature shows aVNS can safely improve cognitive functions such as attention, learning, and memory. It has also been shown useful to manage stress responses. For safe and efficient sUAS/UAV operation, it is essential for pilots to maintain high levels of vigilance and decision-making under pressure. By modulating sympathetic stress and cortical arousal, aVNS can prime cognitive faculties before training, help maintain focus during training and improve stress recovery post-training. Furthermore, aVNS has demonstrated the potential to enhance multitasking and cognitive control. This may help remote pilots during complex sUAS operations by potentially reducing the risk of impulsive decision-making or cognitive errors. This paper advocates for the inclusion of aVNS in remote pilot training programs by proposing that it can provide significant benefits in improving cognitive readiness, skill and knowledge acquisition, as well as operational safety and efficiency. Future research should focus on optimizing aVNS protocols for drone pilots while assessing long-term benefits to industrial safety and workforce readiness in real-world scenarios.

## 1. Introduction

Currently the drone industry is experiencing significant global growth. The operation of aerial drones or Remotely Piloted Aircraft Systems (RPAS) is quickly becoming widespread across many other industries. The increased accessibility and use of RPAS including small unmanned aerial systems (sUAS) and unmanned aerial vehicles (UAVs) has driven the need for changes in policy and aviation regulations [1-4]. Applications of RPAS span construction, cargo transport, inspection, communications, surveying and mapping, agricultural and wildlife management, search, rescue and recovery, consumer electronics and entertainment, surveillance, security, and national defense industries [5-8]; **Figure 1**). There are growing needs for the development and implementation of training programs designed to provide remote pilots (RPs) with the physical and cognitive skills required for safe RPAS operation [1, 9]. The operation of RPAS is a cognitively demanding task that requires distributed attention, efficient decision making, and physical multi-tasking under a high psychological stress load [10]. Developing solutions intended to enhance remote pilot (RP) training and performance represents an immense opportunity for incorporating neurotechnology to improve learning, cognition, and stress responses into the control and command of RPAS.

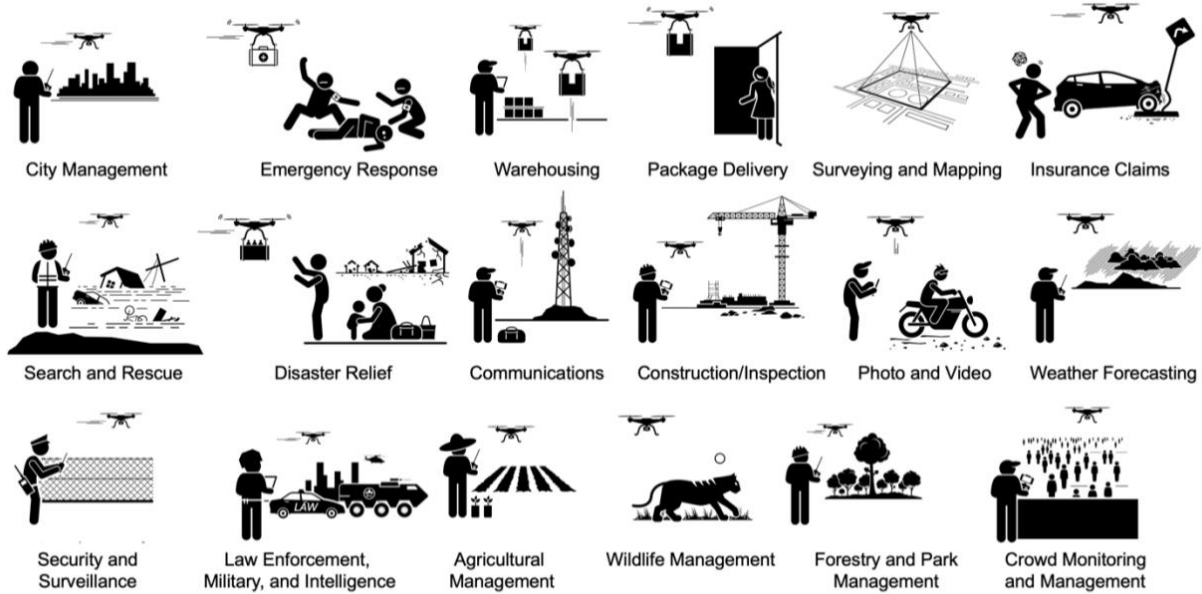

**Figure 1. Commercial and industrial utility of remotely piloted aircraft systems and unmanned aerial vehicles.** The iconographic illustrations depict some example applications of sUAS/UAVs that are creating an impact on many different industries. The development and deployment of different types of sensors and edge computing methods has led to a growing adoption of drones for remote applications. This broadening use of semi-autonomous sUAS/UAVs in commercial, security, and defense applications means workforce and talent development efforts will also grow to include the establishment of RP training programs and certifications. Some training programs will incorporate modern neurotechnology methods, such as auricular vagus nerve stimulation to enhance RP cognition and performance.

Like the growth of drone industry, the field of neuroengineering has also experienced recent growth. There have been many advances in the development and commercialization of methods and devices for sensing and modulating human brain activity and behavior. Some of these approaches can be useful for enhancing RP training and operations. For example, noninvasive methods of recording brain activity like electroencephalography (EEG) and functional near infrared spectroscopy (fNIRS) have demonstrated feasibility for enabling thought-controlled drone flight and drone swarming [11-14]. Similarly, noninvasive neuromodulation methods like transcranial electrical stimulation (tES) and noninvasive vagus nerve stimulation (VNS) have been shown to enhance learning during pilot training, as well as pilot operations in flight simulations [15-17]. As discussed below, integrating these types of noninvasive neurotechnologies into the drone industry can enhance the safety and efficiency of sUAS operations. This article places an emphasis on the use of noninvasive VNS for enhancing RPAS training, cognitive readiness, and operational performance. Advancing noninvasive VNS to enhance RPAS training and performance will provide a unique testbed for designing, prototyping, testing, and validating a next generation brain machine interface (BMI) that can deepen our ability to interact with machines.

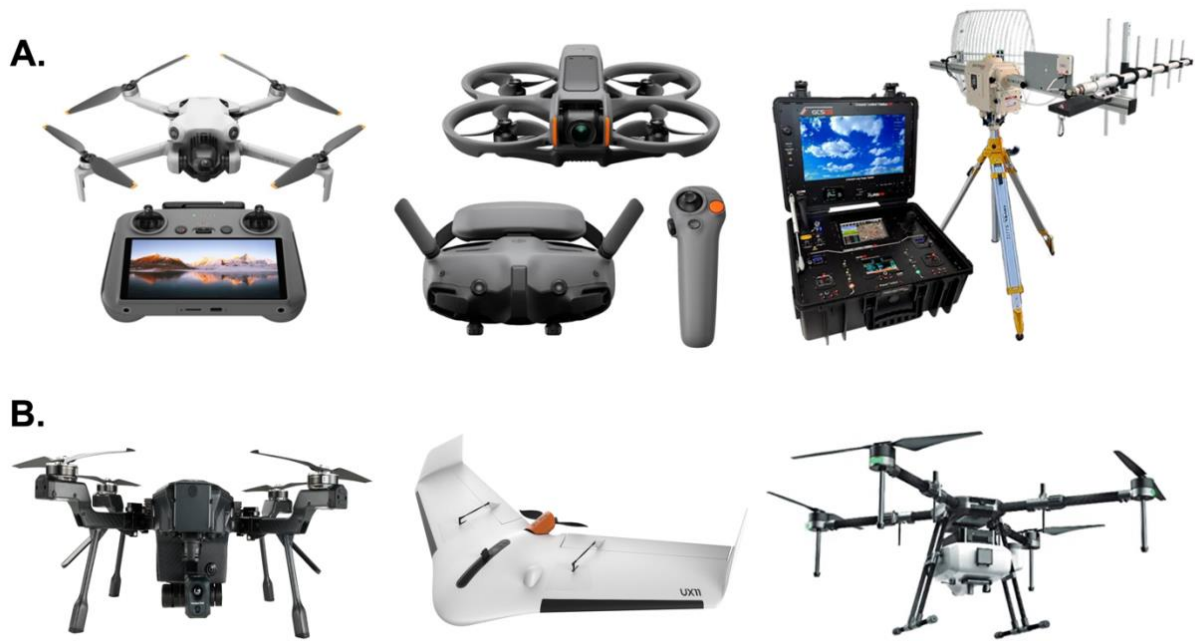

**Figure 2. Diversity of small, remotely piloted aircraft systems and controller interfaces.** Modern aerial drones are equipped with numerous features including obstacle avoidance and flow sensors, gyroscopes and accelerometers, global positioning system (GPS) sensors, many types of optical, infrared, and multi-spectral cameras, as well as other sensors and actuators. **A.** Several different types of drone quadcopters or small unmanned aerial systems (sUAS) are shown in addition to various controller interfaces. The DJI Mavic Mini drone (*left*) is shown with a handheld remote-control system, which includes flight joysticks and a display for GPS information, navigation data, speed, altitude, battery information, and a real-time video feed from a CMOS camera sensor. The DJI Avata 2 (*middle*) is shown with a first-person view (FPV) head-mounted display and a motion sensitive controller. Using the sUAS configuration shown, RPs navigate and control the drone using a combination of motion gestures and buttons while using FPV visual inputs streamed from the drone. The drone ground control system (*right*) shows a more sophisticated mobile control center with a RF antenna for streaming along information including video feeds and sensor data. **B.** Larger versions of sUAS systems are shown to illustrate broad applications. The Teledyne FLIR SIRAS (*left*) is an industrial quadcopter drone capable of thermal imaging for inspection, surveillance, search, and rescue operations. The Delair UX11 fixed-wing drone (*middle*) was designed for professional mapping applications while achieving long flight ranges. The DJI Agras T40 (*right*) is another sUAS quadcopter capable of higher payloads designed for agricultural applications including mapping, inspection, and spraying. The sUAS/UAVs shown in **A** and **B** all require different skills, levels of attention, and cognitive engagement ranging from simple to complex. The drones illustrated represent a small fraction of the types of aerial drones available to industry and consumers.

# 2. Essentials of Remote Pilot Training and Operational Preparedness

All RPs of sUAS/UAVs must acquire some basic technical skills and fundamental knowledge for safe operation. The Federal Aviation Administration (FAA) and other regulatory agencies have established rules and regulations that must be understood and followed by RPs. These regulations including required FAA RP certifications and drone registrations depend upon the size or class of RPAS being operated, purpose or type of flight missions being conducted, and geographical region of operations. RPs must learn and understand the laws and regulations that govern drone operations, including airspace control and restrictions, no-fly zones, weather reports, privacy concerns, and aviation safety. Additionally, there are many types of drones and control interfaces available to RPs, so there are needs to acquire specific skills and knowledge related to the platforms used and operations conducted (**Figure 2**). RPs should also undergo appropriate training and be able to demonstrate proficient knowledge and skills in drone operation (takeoff, navigation, and landing), flight physics and mechanics, drone maintenance and power management, troubleshooting technical issues, and emergency procedures. Training in data management may also be required for RPs to handle the types of data collected by drones, including storage, processing, and interpretation or analysis for specific applications and operations. Modulation of brain activity at different phases of RP training, as discussed below using noninvasive auricular VNS (aVNS), can be used to enhance learning and skill acquisition.

Even though most RPAS have a high degree of automation, their operation still requires significant cognitive resources by the human operator or RP. In fact, the cognitive abilities and skills of drone pilots are critical for safe and efficient sUAS/UAV operation. Drone pilots must maintain accurate perception of the drone's position, orientation, and environment. This can present unique sensory processing challenges to some drone operators. For example, RPs experience mismatches between their body's visual and proprioceptive systems during first person view (FPV) navigation using drone vision [18]. Relying on situational awareness and distributed attention, RPs must learn to make decisions and coordinate physical actions to ensure safe and efficient drone performance. Vigilance or sustained attention must be dedicated to monitoring flight parameters, power levels, and system information to prevent accidents and ensure smooth operations. Rapid and effective decision-making is essential, particularly in emergency situations. Training should focus on enhancing cognitive flexibility and problem-solving abilities. RPs often need to manage multiple tasks simultaneously, such as navigating the drone, monitoring sensors, and communicating with crew members. Since drone operation is a mentally demanding task, evidence that noninvasive VNS can enhance cognitive readiness, attention, and decision-making is discussed in the context of enhanced RP cognition (ERPC).

Tied closely to the cognitive demands of remote piloting discussed above, significant stress can arise during sUAS/UAV operations. Managing stress is paramount to ensuring flight safety, since poor decision-making and human error are leading causes of drone crashes. Training RPs to recognize the signs of stress and understand its impact on mental and physical performance is critical. Further, RPs should be provided with tools and methods to manage stress efficiently, such as breathing exercises, structured rest periods, and noninvasive VNS. It has been shown that noninvasive VNS can safely and rapidly dampen sympathetic nervous system activity [19, 20]. Use of noninvasive VNS in this manner can help RPs manage stress during training and flight operations as further discussed below. Implementing other approaches to help RPs manage stress, such as establishing support networks for pilots, including access to mental health resources and peer support groups is also essential to RPAS aviation safety. Given that fatigue and poor sleep increase psychophysiological stress, programs should also aim to provide RPs with tools and work environments that help manage fatigue. This may include policies establishing mandatory rest and recovery periods, resources for workload management, and implementing noninvasive VNS to improve sleep quality and reduce fatigue. Several studies have shown that noninvasive VNS can improve sleep quality, reduce fatigue, and improve vigilance in ways that would benefit RP stress, cognition, and performance [16, 20-27].

# 3. Auricular Vagus Nerve Stimulation Science and Technology

The activity of the vagus nerve (Cranial Nerve X) underlies core aspects of our health including digestion, cardiovascular reflexes, cardiac activity, immune responses, arousal (sleep/wake, consciousness, and fight/flight), attention, cognition, learning, and memory. There are different methods of vagus nerve stimulation (VNS). Some methods involve surgical procedures to wrap the cervical vagus with metallic electrodes connected to a pulse generator and battery placed underneath the skin as a medical device to treat epilepsy and depression. Some forms of noninvasive cervical VNS (cVNS) use metal electrodes coupled to the skin with an electrolyte gel to conduct transcutaneous pulsed electrical currents to the cervical branch of the vagus nerve. Another noninvasive approach, known as transcutaneous auricular VNS (aVNS) involves the electrical stimulation of auricular branches of the vagus nerve (ABVN; Arnold's nerve or Adleman's nerve) located under the skin's surface of the external ear. There is some conjecture and academic debate regarding where and how vagal fibers should be targeted on the ear. The known physical anatomy of ABVN fibers however enables repeatable and accurate targeting by placing electrodes in the external acoustic meatus, which allows some devices to be used just like ear bud headphones (**Figure 3**).

Branches of the ABVN innervating the external acoustic meatus serve as the functional anatomical basis for Arnold's cough reflex [28-30] and the mammalian diving reflex observed during full facial (ear) submersion [31-35]. This region of the external ear is also innervated by branches of the trigeminal nerve (Cranial Nerve V), so both trigeminal and vagal fibers are stimulated when using this approach. This anatomy further highlights the close functional relationship between the trigeminal nerve and the vagus nerve on the face (around the eyes, nose, and mouth) and ears in natural reflexes like the mammalian diving reflex and trigemino-cardiac reflexes [33, 36-40]. These trigeminal and vagal networks and their afferent connections are central to the ability of aVNS to suppress sympathetic activity and enhance focus.

A cVNS device (GammaCore, Electrocore, Inc) has been cleared by the US Food and Drug Administration to treat headache and an aVNS device (Sparrow Ascent, Spark Biomedical, Inc) has been cleared to treat symptoms related to opiate withdrawal. The aVNS form factor is preferred over cVNS due to its ability to offer hands-free operation, form factor familiarity, and ease of use in targeting vagus nerve fibers (**Figure 4**). There are several aVNS devices on the market available to consumers for lifestyle, health, and wellness applications (**Figure 3A**). Over the past decade, aVNS has gained attention for its demonstrated ability to safely modulate autonomic nervous system activity, inflammation, attention, learning, sleep, and activity in one of the brain's central arousal control regions the locus coeruleus (LC) [25, 41-48]. Human factors elements, design principles, and biomedical engineering considerations of aVNS methods and devices are critical to ensure user comfort, effectiveness, and adoption [41, 49, 50].

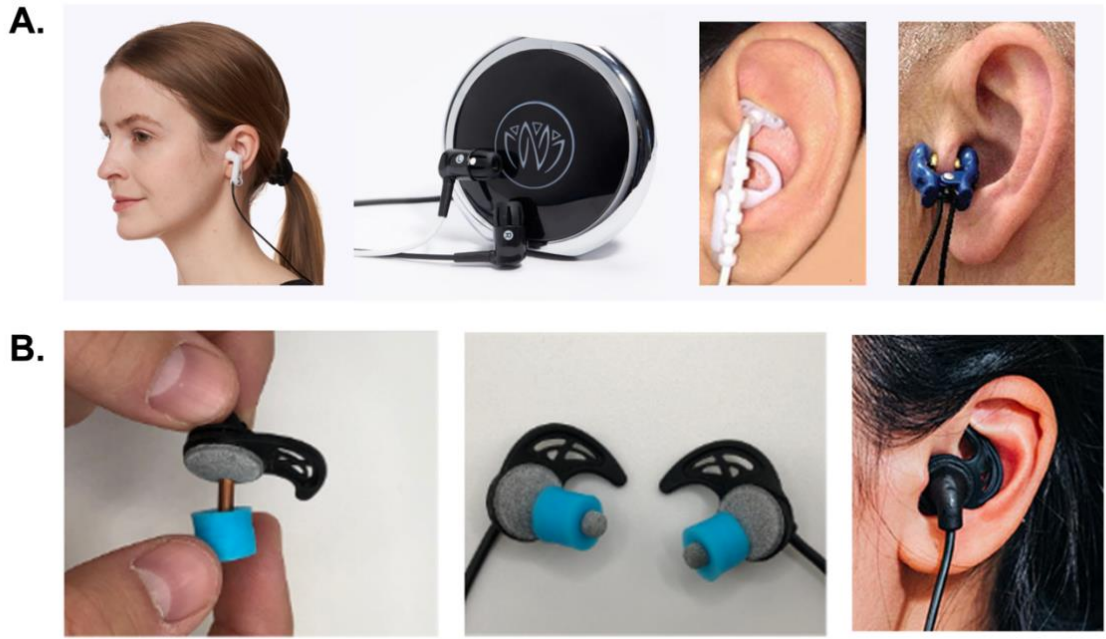

**Figure 3. Methods and approaches used for auricular vagus nerve stimulation.** Several different technical approaches to conducting auricular vagus nerve stimulation (aVNS) have been developed. From human factors and biomedical engineering perspectives, these aVNS methods use different skin-electrode coupling methods. Ensuring user comfort is critical to aVNS efficacy as any distracting or uncomfortable sensations will produced a confounding increase in sympathetic arousal. **A.** Some aVNS methods and devices utilize metal clip electrodes like the clip (Soterix Medical, Inc) shown on the *left*. These clips are used to mechanically couple the skin to a metal electrode using an electrolyte solution or gel. This approach creates a distracting pinching sensation and can produce electrical biting or prickling sensations. The Xen (Neuvana, Inc) aVNS device shown (*middle-left*) implements a different skin-electrode coupling approach using a saline-sprayed conductive rubber electrode placed in the left external acoustic meatus. This creates a distracting wet feeling in the ear of users. Due to body motion, fluid flux, absorption, and dehydration this wet coupling method also leads to electromechanical distortions in the fluid coupling layer between hard charge carrier (i.e. metal electrode) and irregular surfaces of the skin. A problem in general for the entire field of bioelectronics, electromechanical mismatches between the electrode and soft, sensitive skin lead to local electrical hotspots and cause distracting sensations or discomfort for the user. Other aVNS methods like the Nemos device (Cerbomed GmbH) shown *middle-right* use small, steel ball electrodes that can produce high current densities resulting in discomfort or electrical biting and stinging sensations. The Tinnoff device (SaluStim Group) shown at *right* features a different type of aVNS clip electrode that can cause distracting sensations as discussed. **B.** Images of the BRAIN Buds aVNS electrodes (IST, LLC) shown in the *left* and *middle* panels were developed as conductive hydrogel earbud electrodes to produce an easy-to-use, comfortable user-experience. To overcome the electromechanical problems when coupling electrodes to the skin discussed, materials and bioelectronics engineers have recently developed soft hydrogels that mimic the electrical, thermal, and mechanical properties of skin. Using conductive hydrogels to couple electrodes to the skin results in more uniform current distributions and enhanced user comfort during transcutaneous electrical stimulation. As shown on the *right*, BRAIN Buds were designed as a bilateral aVNS system to be used like earbud headphones.

Most existing aVNS methods and devices fall short in providing user comfort due to their reliance on poorly designed approaches of coupling electrodes to the ear (**Figure 3A**). Discomfort during stimulation manifests as an electrical biting, burning, or stinging sensation, primarily due to electromechanical mismatches between the electrode and sensitive skin of the ear. Electrical and mechanical differences between hard charge carriers (i.e., metal electrodes) and the soft, irregular surface of the skin produces asymmetrical current distributions and localized electrical hot spots that can be annoying or distracting. These mismatches are aggravated by wet coupling methods, which use a saline spray or electrolyte gel coating on conductive rubber or metal electrodes. In this case, due to body motion, fluid flux, and dehydration, the microfluidic interface at gaps between the skin and electrode undergoes frequent mechanical distortions affecting local electrical impedances. Saline-sprayed rubber electrodes designed to be placed in the ear as shown in **Figure 3A** also cause wet and other distracting sensations that may reduce VNS efficacy [51]. Methods that use small metal electrode balls or other small area of skin/metal contact can produce high current densities resulting in uncomfortable sensations during stimulation. Metal or rubber electrodes clipped onto the ear at the tragus or other regions create mechanical pinching sensations that are distracting and uncomfortable (**Figure 3A**).

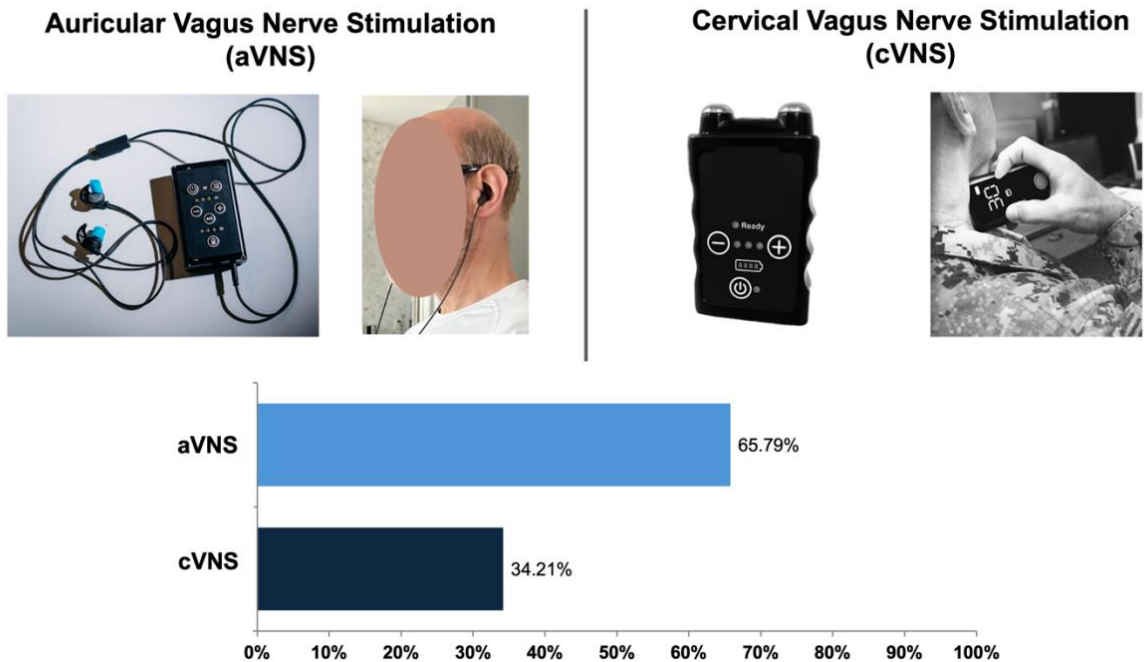

**Figure 4. Auricular vagus nerve stimulation is a preferred approach to noninvasive VNS.** Images showing the BRAIN Buds aVNS system (IST, LLC) on the *left* and the Tac-stim cVNS device (Electrocore, Inc.) on the *right* as shown to potential users (N = 524) in a customer preference survey. Significantly more users indicated they preferred the aVNS device over the cVNS device as illustrated by the histograms (*bottom*). The hands-free and easy-to-use form factor of the aVNS device with its superior hydrogel skin-electrode coupling method make it a preferred device for enhancing RP training and operational performance.

As previously alluded, uncomfortable experiences or distracting sensations caused by wearing or being stimulated by an aVNS method or device can detract from, minimize, or cancel the intended effects on sympathetic arousal (stress), cognition, and neuroplasticity [51, 52]. To improve the comfort, efficacy, and usability of aVNS, a new approach using earbud electrodes made from a soft, skin-like, low impedance, conductive hydrogel was recently developed (**Figure 3B**). These hydrogel earbud electrodes conform to the skin of the acoustic meatus to target trigeminal and ABVN fibers. This approach was specifically engineered to overcome issues, which have plagued aVNS methods and device designs due to electromechanical mismatches between electrodes and the skin of the ear as described above. The use of hydrogel coupling methods reduces these electromechanical mismatches to ensure stable, uniform current distributions across skin-electrode interface resulting in enhanced user comfort and electrical efficiency [53-59]. The use of hydrogel earbud electrodes for aVNS results in a safe, easy to use, comfortable, and effective experience that can provide numerous benefits to RPAS operators and crew members.

# 4. Auricular Vagus Nerve Stimulation for Enhancing Remote Pilot Training

Based on the data from numerous scientific studies, there are several ways in which aVNS can be used just to support and enhance RP training, skill development, and performance (**Figure 5**). The training required to safely and efficiently conduct RPAS operations requires a lot of time, motivation, and learning to acquire necessary skills and knowledge. Suggesting aVNS may be used to improve the motivation of RPs to engage in proper training, it has been shown that aVNS boosts human drive to work for rewards [24]. Several studies have also shown that aVNS can improve human learning and memory [22, 23, 27, 60-64]. In addition to direct effects on neuroplasticity, the influence of aVNS on learning and memory can be also attributed to its ability to modulate human cortical arousal and attention [25, 65-69]. It is well established that reducing stress can also enhance cognition and learning efficiency. Several studies have shown that aVNS can reduce the psychological and neurophysiological symptoms of stress [20, 66, 70-72]. By modulating ascending arousal systems of the brain, the data collectively indicate aVNS can be a useful tool to enhance the attention, learning, and cognitive readiness of RPs (**Figure 6**). This may be achieved by administering aVNS prior to learning or training sessions to prime pilots' cognitive faculties and improve focus or attention. Taking an approach using aVNS in pre-training conditioning paradigms can lead to improved training outcomes, where RPs are better able to absorb and retain critical skills and information.

It is also feasible to use aVNS during training sessions to enhance real-time cognition for improving RP learning and performance. Continuous or intermittent use of aVNS during training can help RPs maintain high levels of cognitive function and focus. For example, using an intermittent stimulation for pilot enhanced cognition and training (InSPECT) approach may be particularly advantageous during complex simulations or when a RP must learn new, challenging maneuvers. In fact, aVNS has been shown to improve action control performance and response selection when task demands are high [73]. It also been demonstrated that aVNS can improve cognitive control during multi-tasking to enhance performance [74]. The ability of aVNS to dampen stress responses is likely a contributing factor to the improved performance observed under high cognitive loads. Another suggestion to improve learning and skill development in training is to use aVNS during emergency flight simulations or in stress inoculation sessions when RPs are intentionally exposed to high-stress scenarios in controlled environments. By modulating activity in the locus coeruleus (LC) and the body's stress response, aVNS can help RPs remain calm and composed, even in high-pressure training scenarios (**Figure 6B**). As discussed further below, this reduced stress can improve the ability of RPs to make clear, rational decisions and perform well under pressure. The reduction of stress by aVNS also suggest it can be used following training sessions to improve rest and recovery from mental strain [75]. Since most studies to date have incorporated aVNS before or during tasks, future research should aim to determine whether aVNS can be used following training sessions to enhance memory consolidation and improve skill/knowledge retention.

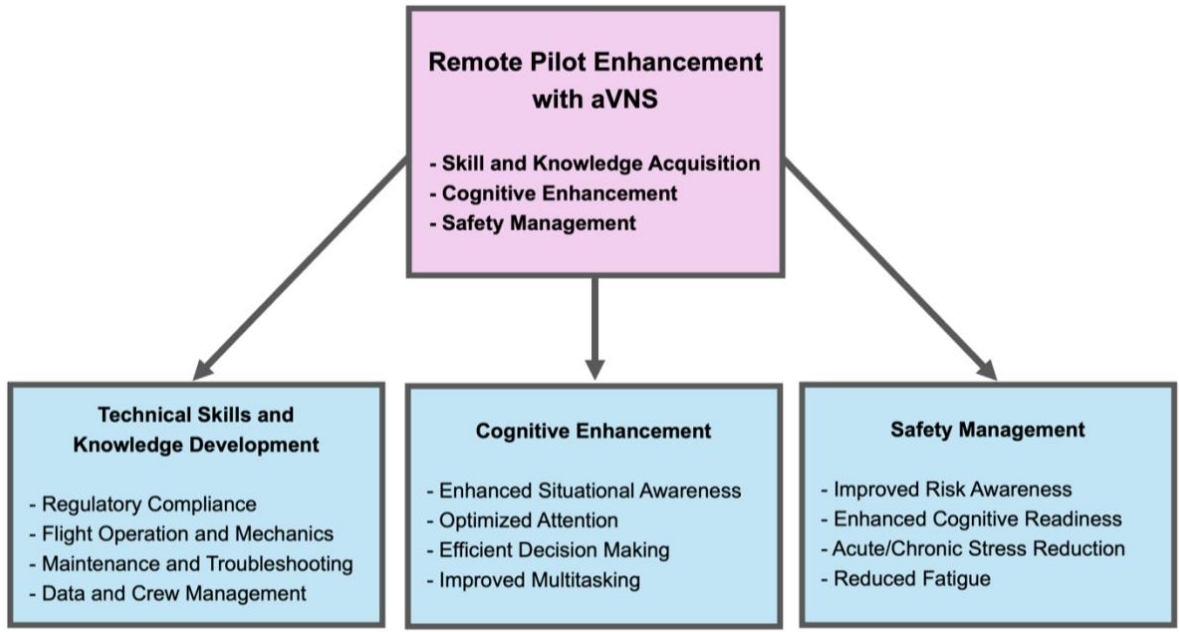

**Figure 5. Influence of auricular vagus nerve stimulation on RP training and performance.** The diagram illustrates how auricular vagus nerve stimulation can enhance RP performance by improving skill and knowledge acquisition and retention, improving cognitive abilities, and improving operational safety.

# 5. Auricular Vagus Nerve Stimulation to Improve Remote Pilot Cognitive Control, Multitasking, and Decision Making

Multitasking and cognitive control is critical to safety and operational success during sUAS/UAV flight, when pilots must navigate the drone, monitor power, navigation, and sensor systems, and communicate with crew members. Several methods and approaches can be taken to conduct aVNS during RPAS operation, including during FPV sUAS operation to enhance cognitive control and multitasking performance (**Figure 7**). Working memory and cognitive control networks are critical in decision making [76-79]. Impaired working memory has been shown to underlie impulsivity in decision making [80]. Impulsive decision-making by RPs presents risks to RPAS operations since it can lead to flight emergencies or crashes injuring people and property. Thus, improving RP decision-making and reducing impulsivity during RPAS operation would be a desirable outcome for aVNS-modulated human drone interactions (**Figure 6A**). It has been demonstrated that aVNS can improve human working memory [81] and cognitive flexibility [82]. Other studies have shown that aVNS can produce more efficient neural processing requiring fewer resources to achieve cognitive control [83], as well as to improve cognitive control or adaptation in response to conflict [84]. As cited above, aVNS improves action control performance when task demands are high [73] and enhances cognitive control during multi-tasking [74]. These data indicate aVNS may improve cognitive control and flexibility, enabling pilots to switch between tasks and manage multiple streams of information more efficiently. By enhancing executive function and reducing the influence of high cognitive loads, aVNS can improve RPs' decision-making abilities. This may help ensure that RPs can make quick, accurate decisions in dynamic operational environments.

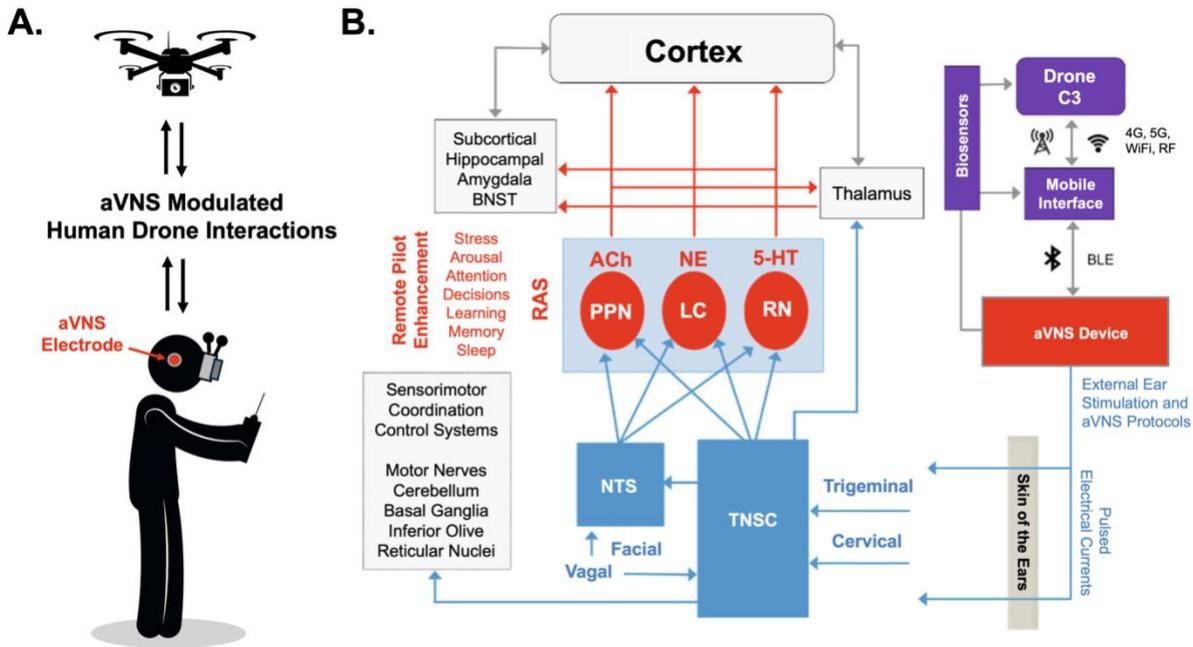

**Figure 6. Networks and neurocircuitry for enhancing remote pilot performance with auricular vagus nerve stimulation. A.** The schematic illustrates the use of transdermal or transcutaneous auricular vagus nerve stimulation (aVNS) to enhance human drone interactions and RP performance. **B.** aVNS enables bottom-up modulation of higher brain regions by influencing nuclei of the ascending reticular activating system. The schematic illustration shows brain circuits and general concepts involved in the bottom-up regulation of neural networks to enhance plasticity and performance using aVNS. The delivery of pulsed electrical currents across the skin of the ear can modulate the activity of cranial and cervical nerves (blue pathways). The trigeminal, cervical, facial, and vagus nerves send afferent inputs to the trigeminal nucleus sensory complex (TNSC) and nucleus of the solitary tract (NTS) in the brain stem. These first nuclei then send projections to higher nuclei of the ascending reticular activating system (RAS). The RAS includes pedunculopontine nuclei (PPN), the locus coeruleus (LC), and raphe nuclei (RN), which transmit acetylcholine (ACh), norepinephrine (NE), and serotonin (5-HT) respectively to various parts of the brain. These regions and processes highlighted in red serve as the chief control centers in the brain for regulating arousal, attention, learning, memory, sleep onset, and stages of sleep. Changes in these networks and brain regions produced by aVNS can enhance RP training and performance by improving stress responses, attention, learning, memory, cognitive control, multitasking, and decision making. In the future, biosensors measuring attention, cognitive load, and stress may be used in a closed-loop manner with aVNS as part of an integrated drone command, control, and communications platform to enhance RP cognition and performance in real-time.

As previously described, the mammalian diving reflex was designed to optimize autonomic activity in response to auricular vagus (ear) and trigeminal (ear, face, mouth, and nose) nerve stimulation during full facial submersion in water [31-35]. This reflex is well documented to result in a decreased heart rate and oxygen conservation in the face of a high stress manipulation. The mammalian diving reflex induces a paradoxically relaxed state, due to natural vagus nerve stimulation during breath-hold diving and other underwater activities involving breath holding [85]. Breath-hold diving, or free diving, can produce unique neurophysiological sates associated with a high degree of focus, attention, and cognitive control. Thus, it is not shocking that aVNS conducted with pulsed electrical stimuli can produce similar types of cognitive and psychophysiological effects. Studies have indeed shown aVNS can reduce sympathetic activity and stress [20, 25, 86-90]. Operationally, using aVNS before RPAS operations may lower anxiety and stress levels, allowing RPs to begin their missions with a calm and focused mindset. This proactive approach may enhance flight safety by reducing stressed-induced cognitive errors or failures. Another approach is to utilize aVNS during drone operations, including sUAS FPV flight (**Figure 7**), to help RPs manage in-flight stress, particularly during high-stakes or emergency operations. By stabilizing the autonomic nervous system, aVNS may reduce RP stress from impairing multitasking performance and decision-making. Additional research and development of aVNS to support RP training and performance will continue to expand upon the basic approaches described above.

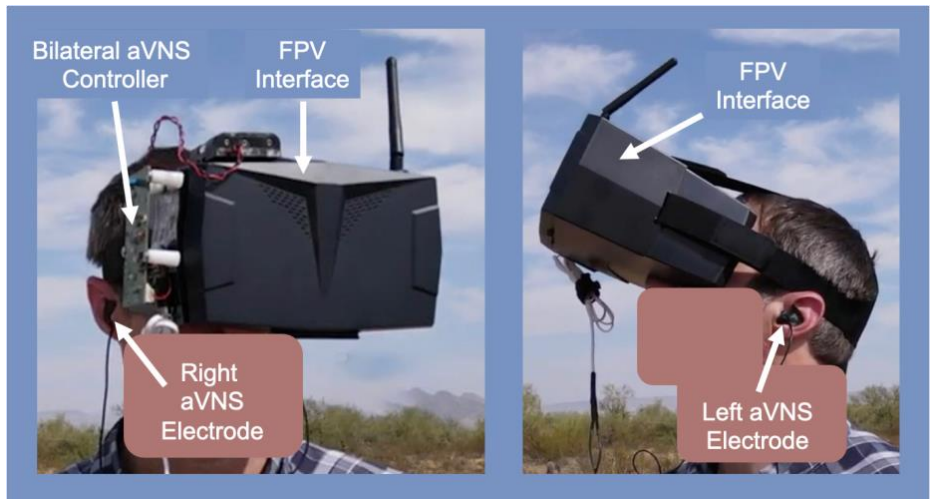

**Figure 7. Auricular vagus nerve stimulation for real-time optimization of FPV sUAS remote pilot performance.** The photographs illustrate a functional aVNS/FPV display headset prototype (IST, LLC). The prototype was used to deliver bilateral aVNS through hydrogel earbud electrodes during FPV sUAS flight training. The photographs show the aVNS controller mounted to the FPV display interface so RPs can safely and easily activate a short aVNS session anytime during the FPV flight.

## 6. Future Considerations: Closed-loop aVNS BCIs for enhancing remote pilot performance

Bridging neurotechnology with drone innovation will encourage the development of new types of closed-loop, brain-computer interfaces (BCIs) intended to improve human-computer interactions (HCIs). BCI studies have shown that brain signals recorded from EEG and fNIRS sensors can be used to directly control drone flight and behavior [11-14, 91-93]. Several near-term opportunities for enhancing RP training and performance with existing aVNS methods and devices were described above. For FPV sUAS operations, neurotechnology hardware for aVNS is readily compatible with existing heads up displays (**Figure 7**). There are other efforts to develop noninvasive, closed-loop brain sensing and brain neuromodulation devices (i.e., multimodal neural interfaces) to monitor and enhance cognition and attention [47, 94]. Integrating electroencephalographic (EEG) or functional near infrared spectroscopy (fNIRS) sensors into existing FPV display designs can be accomplished to monitor cognitive load, stress, or attention. Similar approaches have been used to monitor the psychophysiological and cognitive states of drone pilots [13, 95, 96]. Some proposed aVNS-EEG closed-loop systems for modulating attention can be designed to fit in the ear [94]. Such designs will be advantageous to RPs due to their ease of use. Soon there will be several growing efforts to engineer a new generation of BCI-FPV display controllers for providing real-time enhancement of drone pilot cognition and performance during dynamic flight operations.

Another approach to engineering closed-loop BCIs may be to connect the drone control, command, and communications (C3) modules with aVNS hardware to enhance pilot skill training. As mentioned before an InSPECT approach can be taken where intermittent aVNS stimuli provide a training or reward cues for skill training (**Figure 6 and 7**). Data from drone position and navigational sensors can be used to trigger aVNS stimuli prior to a skill training command or as a reward following a successful maneuver or skill demonstration to reinforce learning and development. In support of these approaches, it has been shown that aVNS can significantly improve motor action planning, enhance motor sequence learning, and improve associated motor cortex efficiency [97, 98]. Auricular VNS also boosts human motivation to work for reward [24]. In stroke patients, task pairing with aVNS improved fNIRS signals in motor cortex, increased the frequency of motor evoked potentials, and improved motor recovery [99]. Closed-loop electromyography-triggered aVNS has also been used in motor training and rehabilitation to improve motor recovery following stroke [100]. These data demonstrate the ability of aVNS to trigger neuroplasticity in reinforcing motor and task learning in healthy humans and following neurological injury. It therefore seems reasonable to hypothesize aVNS can also be used in a variety of closed-loop manners to reinforce cognitive and motor learning required for RPs to operate drones in complex situations.

The basic approaches discussed above can be tested and validated for RPs while being evaluated for broader applications. The approaches discussed may be useful for solutions intended to improve HCIs in the larger robotics industry. It should be investigated whether aVNS can enhance other human-robot interactions, where we cooperate with machines using specific skills and knowledge to accomplish large or complex tasks in manufacturing, construction, medicine, shipping, transportation and delivery, city and vehicle maintenance, deep-sea and space exploration, and other areas.

## 7. Conclusions

The integration of aVNS into remote pilot training and operations represents a promising approach to enhancing cognitive readiness, stress management, and overall operational performance in the rapidly growing drone industry. The scientific evidence underscores the effectiveness of aVNS in improving cognitive functions such as attention, learning, and memory, while also offering significant benefits in reducing stress and enhancing multitasking abilities. These enhancements are crucial for remote pilots who face complex and demanding operational environments that require high levels of vigilance and rapid decision-making. By incorporating aVNS into training programs, we can create more resilient, focused, and efficient remote pilots, ultimately improving the safety and efficacy of sUAS/UAV operations.

Looking forward, the application of aVNS as part of a broader neurotechnology strategy has the potential to transform the field of remote piloting. Future research should continue to refine aVNS protocols and explore their integration with other neurotechnologies, such as BCI, to develop closed-loop systems that provide real-time cognitive enhancement during drone operations. Additionally, as these technologies evolve, it will be essential to validate their long-term benefits in real-world scenarios, ensuring that they contribute not only to individual pilot performance but also to broader industrial safety and workforce readiness. Through continued innovation and research, aVNS and related neurotechnologies can help establish new standards for training and operational excellence in the drone and robotics industries.

**Acknowledgements**
The development of BRAIN Buds by IST, LLC was funded based on research sponsored by Air Force Research Laboratory under agreement number FA8650-18-2-5402. The U.S. Government is authorized to reproduce and distribute reprints for Government purposes notwithstanding any copyright notation thereon. The views and conclusions contained herein are those of the authors and should not be interpreted as necessarily representing the official policies or endorsements, either expressed or implied, of Air Force Research Laboratory (AFRL) or the U.S. Government.

**Disclosures**
WJT is a co-founder and equity holding member of IST, LLC. WJT has several pending and issued patents related to the methods of neuromodulation described for enhancing cognition, skill training, learning, and human performance, including semi-autonomous vehicle operation.